\documentclass[aps,prl,twocolumn,floatfix,superscriptaddress]{revtex4}
\usepackage{graphicx,amsmath,amsfonts}
\usepackage{textcomp}
\usepackage{units}

\newcommand{\nm}[1]{\unit[#1]{nm}}
\newcommand{\mm}[1]{\unit[#1]{mm}}

\newcommand{\GHz}[1]{\unit[#1]{GHz}}
\newcommand{\MHz}[1]{\unit[#1]{MHz}}
\newcommand{\kHz}[1]{\unit[#1]{kHz}}
\newcommand{\V}[1]{\unit[#1]{V}}

\newcommand{\ybodd}{$^{171}$Yb$^+$}
\newcommand{\ybeven}{$^{172}$Yb$^+$}
\newcommand{\yb}{Yb$^+$\,}
\newcommand{\level}[1]{$|$#1$>\,$}
\newcommand{\Shalf}{S$_{1/2}$}
\newcommand{\Phalf}{P$_{1/2}$}
\newcommand{\Dhalf}{D$_{3/2}$}

\newcommand{\one}{\level{\,S$_{1/2}$,\,F=1\,}}
\newcommand{\zero}{\level{\,S$_{1/2}$,\,F=0\,}}
\newcommand{\Pone}{\level{\,P$_{1/2}$,\,F=1\,}}
\newcommand{\Pzero}{\level{\,P$_{1/2}$,\,F=0\,}}

\begin{document}
\graphicspath{{Images/}}

\title{Electrodynamically trapped \yb ions for quantum information
  processing}

\author{Chr. Balzer}
\affiliation{Fachbereich Physik, Universit\"at Siegen, 57068 Siegen,
Germany}
\author{A. Braun}
\affiliation{Fachbereich Physik, Universit\"at Siegen, 57068 Siegen,
Germany}
\author{T. Hannemann}
\affiliation{Fachbereich Physik, Universit\"at Siegen, 57068 Siegen,
Germany}
\author{Chr. Paape}
\affiliation{Institut f\"ur Laser-Physik, Universit\"at Hamburg,
Luruper Chaussee 149, 22761 Hamburg, Germany}
\author{M. Ettler}
\affiliation{Institut f\"ur Laser-Physik, Universit\"at Hamburg,
Luruper Chaussee 149, 22761 Hamburg, Germany}
\author{W. Neuhauser}
\affiliation{Institut f\"ur Laser-Physik, Universit\"at Hamburg,
Luruper Chaussee 149, 22761 Hamburg, Germany}
\author{Chr. Wunderlich}
\affiliation{Fachbereich Physik, Universit\"at Siegen, 57068 Siegen,
Germany} \email{wunderlich@physik.uni-siegen.de}

\date{\today}

\begin{abstract}

Highly efficient, nearly deterministic, and isotope selective
generation of Yb$^+$ ions by 1- and 2-color photoionization is
demonstrated. State preparation and state selective detection of
hyperfine states in \ybodd is investigated in order to optimize the
purity of the prepared state and to time-optimize the detection
process. Linear laser cooled Yb$^+$ ion crystals ions confined in a
Paul trap are demonstrated. Advantageous features of different
previous ion trap experiments are combined while at the same time
the number of possible error sources is reduced by using a
comparatively simple experimental apparatus. This opens a new path
towards quantum state manipulation of individual trapped ions, and
in particular, to scalable quantum computing.

\end{abstract}


\maketitle

When investigating fundamental questions related to quantum
mechanics experiments are called for where individual quantum
systems can be accessed and deterministically manipulated. The
interaction of trapped atomic ions among themselves and with their
environment can be controlled to a high degree of accuracy, and thus
allows for the preparation of well defined quantum states of the
ions' internal and motional degrees of freedom. Trapped ions have
proven to be well suited for a multitude of investigations, for
instance, into entanglement, decoherence, and quantum information
processing, and for applications such as atomic frequency standards.
Quantum information processing, in particular, requires accurate and
precise control of internal and often also of motional quantum
dynamics of a collection of trapped ions. In order to eliminate
sources of possible errors, and thus prepare the ground to attain
the ambitious goal of using trapped ions for large scale quantum
computing or quantum simulations, it is desirable to simplify the
apparatus used for such experiments as far as possible.

An unprecedented degree of control of quantum systems has been
reached in recent experiments with trapped ions, for instance, with
Be$^+$ \cite{Leibfried05}, Ca$^+$ \cite{Haeffner05} and Cd$^+$
\cite{Blinov04} ions. Mainly the type of ion used in such
experiments determines the experimental infrastructure needed for
controlled manipulation of these ions. The available ionic
transitions, for instance, determine the radiation sources to be
used: In Ca$^+$ an optical electric quadrupole transition has been
used as a qubit leading to a coherence time limited ultimately by
spontaneous radiative decay. More importantly, phase fluctuations of
the laser light driving the qubit transition limit the available
coherence time, even when using a highly sophisticated light source
\cite{Roos99}. Phase fluctuations of the radiation driving the qubit
transition do not present a major obstacle, if a hyperfine
transition is used as a qubit (as, for instance, in Be$^+$ or
Cd$^+$), since such a transition is usually excited by a stimulated
two-photon Raman process where only relative fluctuations between
the two driving fields limit the available coherence time. Choosing
magnetic field insensitive states as a qubit, as was demonstrated
recently with Be$^+$, may further contribute to achieving the
desired long coherence times \cite{Langer05}.

Another important characteristic that determines the suitability of
a particular ion for experiments requiring quantum dynamics with
minimal error is the initial preparation in one of the qubit states
(before coherent operations take place), and state selective
detection. This usually makes additional light fields necessary,
such that up to a total of seven different light sources are in use
in some experiments \cite{Lucas04}.

For laser cooling of trapped ions a suitable optical electric dipole
transition is usually used. In Ca$^+$ the required wavelengths are
accessible by diode lasers as opposed to Be$^+$ and Cd$^+$ where the
necessary wavelengths are in the deep uv region making more complex
laser systems and non-standard optical elements necessary.

The efficient production of singly charged ions by photoionization
of its neutral precursor was recently demonstrated with Ca$^+$
\cite{Kjaergaard00,Lucas04}. This way of producing ions avoids
difficulties that arise when employing ionization by electron
collision as is done in experiments with Be$^+$ and Cd$^+$ and, at
the same time, allows for isotope selective loading of ion traps.

Spectroscopic studies of Yb$^+$ ions have been carried out in order
to use single ions or an ensemble of these ions to implement
improved frequency standards \cite{Blatt82}. In this Letter we
report on experiments with electrodynamically trapped \yb ions where
advantageous features of previous experiments with trapped Ca$^+$,
Be$^+$, and Cd$^+$ ions are combined: i) Only two light fields,
easily generated by standard laser sources, are needed here for
laser cooling and fluorescence detection \cite{Wunderlich03}. ii)
Isotope selective photoionisation of Yb and nearly deterministic
loading of \yb ions one by one into a linear electrodynamic trap is
demonstrated for the first time. Highly efficient photoionisation is
achieved using just one additional light field delivered by a
readily available diode laser. iii) Laser cooled linear crystals of
individually resolved \yb ions are formed in an electrodynamic trap,
to our knowledge, for the first time.  A laser cooled Yb$^+$ ion
crystal will be useful not only for QIP but also to enhance the
precision of a frequency standard based on ionic optical or
microwave transitions \cite{Bollinger96}. iv) Two hyperfine states
may serve as a qubit in \ybodd\, thus essentially eliminating
spontaneous decay.  Using an optical Raman transition or microwave
radiation \cite{Mintert01} to drive a hyperfine transition allows
for the extension of the coherence time even far beyond a second
\cite{Wunderlich03,Langer05}. v) Investigations of state preparation
and state selective detection are reported and these two processes
are nearly optimized.

An electrodynamic Paul trap with four parallel rod electrodes
(diameter of \mm{0.5}) in a linear quadrupole configuration (inner
radius of \mm{0.75}) is used for rf-confinement of \yb ion crystals
in the radial direction \cite{Maleki89}. Axial confinement is
achieved by applying a DC-voltage to two endcap electrodes (diameter
of \mm{0.4}) spaced \mm{4} apart on-axis, centered between the four
rod electrodes. All electrodes are made of Molybdenum and held in
place by ceramic spacers. The trap is operated at \MHz{21.6} with an
RF-amplitude of approximately \V{400}, resulting in radial secular
frequencies of \kHz{350-450} and an axial secular frequency of
\kHz{40-60}. Alternatively, an RF-drive at \MHz{10.2} is in use
yielding radial and axial secular frequencies of \kHz{800} and
\kHz{65}, respectively. The endcap voltage is typically kept at
about \V{1.0}, in contrast to other linear traps, which report
endcap voltages up to several hundred volts.

For some experiments a miniature Paul trap was used consisting of a
ring electrode of diameter \mm{2} and two endcap electrodes spaced
\mm{$\approx \sqrt{2}$} apart.  The ring trap is operated at
\MHz{9.5} with an RF-ampltiude of \V{700} resulting in secular
frequencies in the x, y and z-directions of approximately \MHz{1.5}.

For each of the traps a photomultiplier is used to detect resonance
fluorescence. In addition, an image intensified CCD (ICCD) camera is
employed to spatially resolve the detected fluorescence. Imaging
onto the ICCD camera is realized with a Questar long-distance
microscope, thus ensuring minimal aberrations and yielding a
magnification of about 30.

\begin{figure}[htbp]
\begin{center}
\includegraphics[width=6cm]{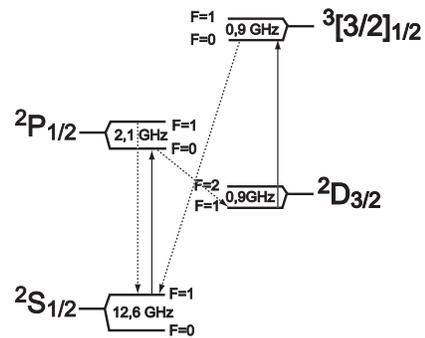}
\caption{Relevant energy levels and transitions of Yb$^+$ including
the hyperfine splitting in \ybodd (drawing is not to scale). The
S$_{1/2}$ - S$_{1/2}$ transition wavelength is 369 nm, and for
D$_{3/2}$ - $[3/2]_{1/2}$ it is 935 nm.}
 \label{fig:Yb}
\end{center}
\end{figure}

Fig. \ref{fig:Yb} depicts relevant energy levels for Yb$^+$ ions.
The electric dipole transition between the \Shalf\, ground state and
\Phalf\, excited state (natural linewidth of \MHz{19.6}) serves for
Doppler cooling, state preparation, and state selective detection
(described below). This transition is driven using laser light with
a wavelength near \nm{369} generated by frequency-doubling the
output of a commercial Ti:Sapphire laser. Optical pumping into the
\Dhalf\, state is avoided by employing a diode laser emitting light
near \nm{935}. The polarization of the light fields is at 45$^\circ$
with respect to the external magnetic field in order to couple all
Zeeman sublevels to the cooling cycle. The \ybodd\, isotope has a
nuclear angular momentum $I=1/2$ that leads to hyperfine-splitting
of all levels.

Several advantages make photoionization the preferred choice to load
single ions into an ion trap. First, it avoids the charging up of
electrically isolating elements near the trap center as with
electron impact ionization. Second, due to its high efficiency and
well-defined interaction volume it allows for a reduction of the
atom flux by several orders of magnitude, thereby minimizing
contamination of the trap electrodes by neutral atoms. Both effects
would lead to a disturbance of the trapping potential. Additionally,
photoionization is an isotope-selective process, which is
advantageous if isotope-selected material is not available, or if
one seeks to trap an isotope of low abundance.

\begin{figure}[htbp]
\includegraphics[width=9cm]{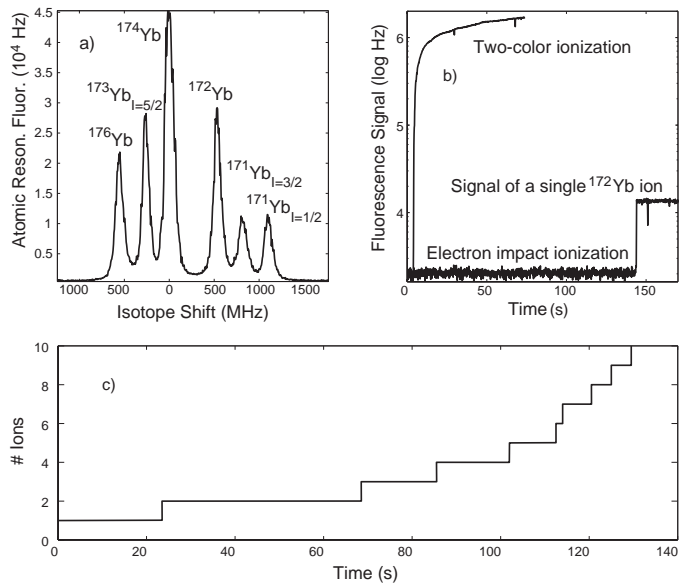}
\caption{a) Isotope selective excitation of the
  6s$_2$ $^1$S$_0$ $\rightarrow$ 6s6p $^1$P$_1$ resonance in Yb (natural
  abundance).  b) Comparison of trap loading rate using electron
  impact or photoionization. c) Number of trapped \ybeven\, ions as a function of time
  demonstrating the possibility of nearly deterministic loading of
  ions.}
 \label{fig:Ionization}
\end{figure}

A diode laser delivering light near \nm{399} with an intensity of
typically \unit[3]{W/cm$^2$} is employed to resonantly excite the
transition between the ground state 6s$_2$ $^1$S$_0$ and the excited
state 6s6p $^1$P$_1$ in neutral Yb. This one-photon process is
isotope selective.  The absorption of a second photon near \nm{369}
(the wavelength used to drive the \Shalf $\leftrightarrow$ \Phalf\,
transition in \yb) leads to ionization (2-color ionization).
Alternatively, the second step of this ionization process is
achieved by absorption of another photon near \nm{399} (1-color
ionization). This is possible since in the presence of the
quasistatic electric trapping field the ionization threshold is
lowered. Both photoionization schemes yield ionization rates in the
experiments reported here that are two (1-color) or three (2-color)
orders of magnitude larger than ionization by electron collision
(keeping the same neutral atom flux), are isotope selective, and
allow for nearly deterministic loading of ion traps.

Fig. \ref{fig:Ionization} depicts important features of the
photoionization process. The collected resonance fluorescence as a
function of wavelength of the laser light exciting the atomic 6s$_2$
$^1$S$_0$ $\leftrightarrow$ 6s6p $^1$P$_1$ transition is shown in
Fig. \ref{fig:Ionization} a). In order to reduce Doppler broadening,
the atomic beam and the laser beam are set at a relative angle of
$90^\circ$, thus allowing to resolve the isotopes with mass 171,
172, 173, 174, and 176 (in a.m.u.). Fig. \ref{fig:Ionization} b)
compares the loading rates for electron impact and photoionization.
It shows the total fluorescence signal as a function of time while
the trap is being loaded. During the time it takes to load a single
ion using electron impact ionization the trap has already been
loaded using photoionization such that the fluorescence signal
saturates (due to the limited acceptance angle of the optical
elements). Here, the rate for electron impact ionization is of the
order $1/150 \approx 0.0067$ \unit{ions/s} while with
photoionization loading rates about three orders of magnitude
larger, approx. \unit[10]{ions/s}, are achieved. Finally, in Fig.
\ref{fig:Ionization} c) the atom flux is reduced such that nearly
deterministic loading of a desired number of ions becomes possible.
The exact number of ions in the trap is determined by counting them
on the spatially resolved fluorescence image of the ICCD-camera. The
loading process can be interrupted at any time by blocking the
ionization laser.

\begin{figure}[htbp]
  \begin{center}
    \includegraphics[width=4cm]{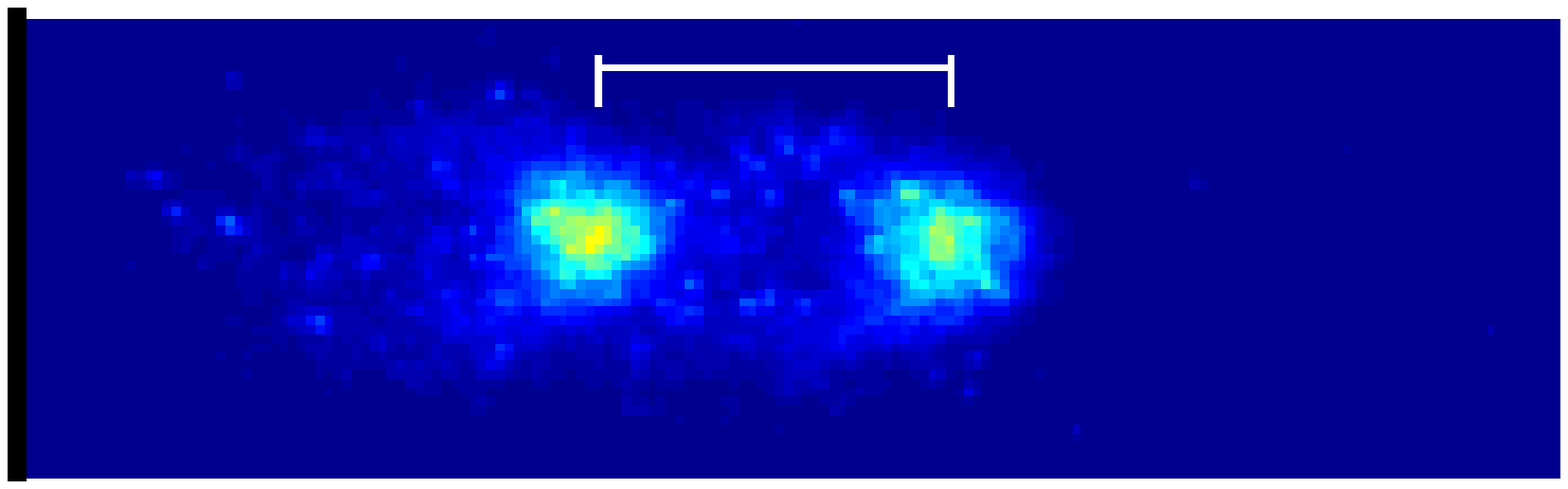}
    \includegraphics[width=4cm]{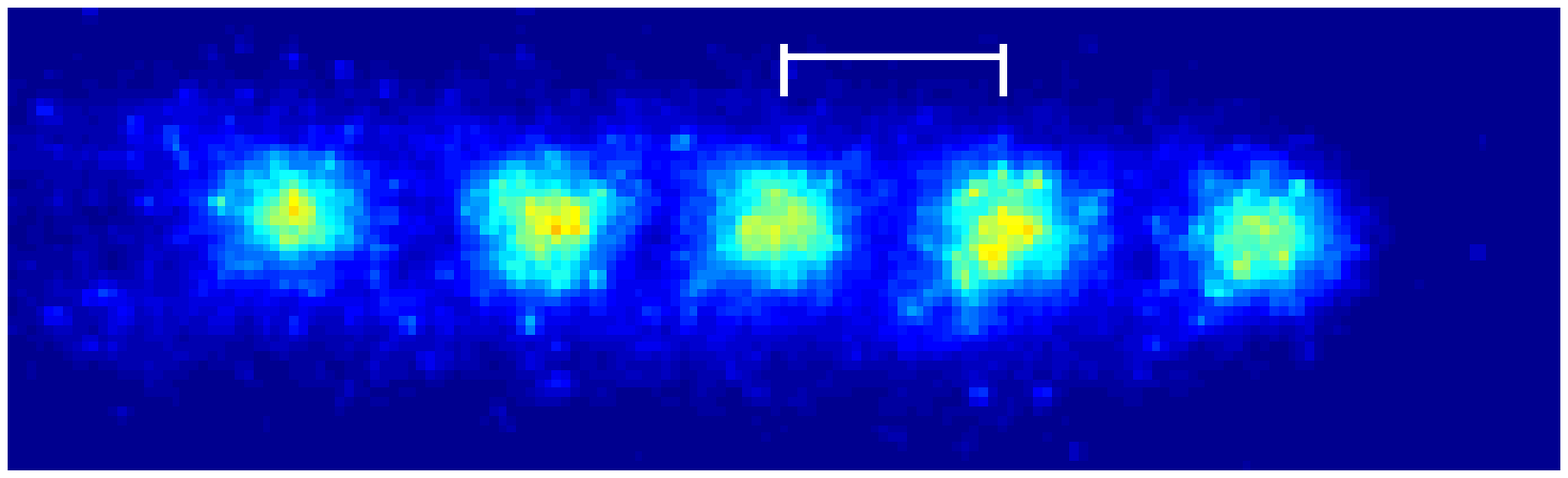}
    \caption{Spatially resolved detection of resonance fluorescence near 369 nm of
    a laser cooled crystal of 2 and 5 \ybeven\, ions, respectively.
    The length scales correspond to 30(2) $\mu$m (left) and 19(2) $\mu$m (right).}
    \label{fig:YbCrystal}
  \end{center}
\end{figure}

Figure \ref{fig:YbCrystal} shows two spatially resolved images of
resonance fluorescence near \nm{369} scattered by a Doppler cooled
crystal of 2 or 5 \ybeven\ ions, respectively (with the fluorescence
intensity color-coded). The images are averaged over 10 frames, with
an illumination time of \unit[200]{ms} for each frame. These images
were recorded with an axial secular frequency $\omega_z = 2\pi \cdot
52$ \unit{kHz}.

A quantum logic operation consists of three steps: first, the qubit
is initialized in a given state. Second, qubit states are coherently
manipulated, and third the resulting state is measured. In the
paragraphs to follow the first and third step for the case of the
\ybodd ion are described. The ability to coherently control the
\ybodd qubit with microwave radiation has been demonstrated
\cite{Hannemann02,Wunderlich03}, and therefore, is not treated here.

\begin{figure}[htbp]
  \centering
  \includegraphics[width=5cm]{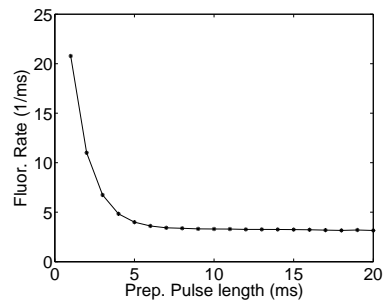}
  \caption{Detected resonance fluorescence near 369 nm from a collection of ions
  as a function of time. The decrease with time indicates optical pumping
  into state \level{0} $\equiv$ \zero. This optical
    pumping process serves for state read-out and initialization (see text).}
  \label{fig:opticalPumping}
\end{figure}

Two hyperfine levels, \level{0} $\equiv$ \zero and \level{1} $\equiv$
\one of the \Shalf\, ground-state of \ybodd serve as a qubit. In order
to keep the experimental setup as simple as possible, it is desirable
to achieve state preparation (as well as state selective detection)
without adding more light sources to the setup. We therefore use the
same optical transition near 369 nm from \one to \Pzero that is used
for laser cooling to also prepare and detect the qubit state. Optical
pumping for state preparation from \one into \zero is achieved by
non-resonantly scattering light near 369 nm off the state \Pone. For
laser cooling, optical pumping is not desired and is hindered by
irradiating the ion simultaneously with radiation at \GHz{12.64}
driving the qubit transition from \zero to \one.
%Without microwave radiation the ion is thus prepared in the
%\level{S$_{1/2}$, F = 0} state by optical pumping.
Fig. \ref{fig:opticalPumping} shows the rate of detected photons
near 369 nm as a function of time after turning off the microwave
radiation, and thus depicts the optical pumping process into state
\level{0}.

The process just described serves simultaneously for state-selective
detection as well as for state preparation, thus initializing the
qubit for subsequent quantum logic operations: If, in step three of
a quantum logic operation the ion is in state \level{0}, no
scattered photons are detected; on the other hand, if the ion is in
state \level{1} a number of photons is detected (see Fig.
\ref{fig:opticalPumping}).

Even though the incident radiation at \nm{369} is detuned by about
\GHz{15} from the resonance \zero $\leftrightarrow$ \Pone, this
non-resonant scattering process leads to depletion of the population
in \zero. As this process competes with state preparation it
diminishes the preparation efficiency. The optical pumping process
has been investigated in detail i) experimentally, and ii), by
solving the optical Bloch equations for this 8-level system in order
to answer the following question: When using just one light field
near 369 nm, and by including the possibility of switching its
polarization and intensity, what is the optimal detection and
preparation efficiency that can be obtained?

\begin{figure}[htbp]
  \centering
  \includegraphics[width=9cm]{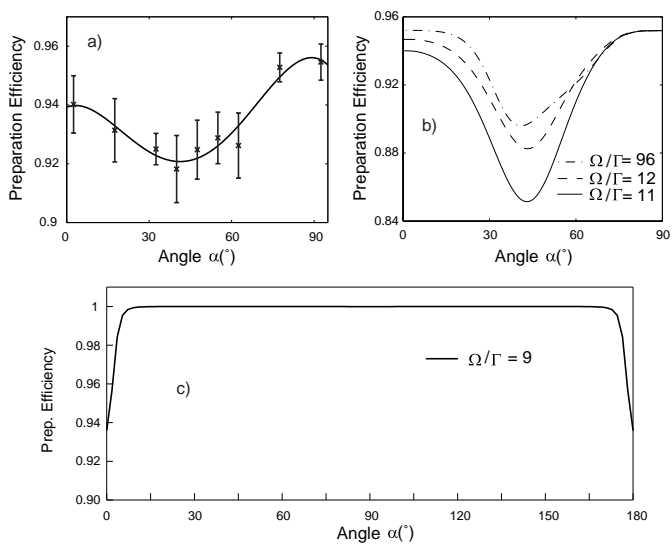}
  \vspace{-1cm}
  \caption{The efficiency of preparation of state \level{0} $\equiv$ \zero
  by optical pumping as a function of the angle $\alpha$ between
  the directions of light polarization and magnetic field,
  respectively. a) Experiment. b) Numerical simulation
  for different ratios of Rabi-frequency and spontaneous decay rate.
  c) Simulation for resonant excitation of \one $\leftrightarrow$ \Pone.}
  \label{fig:preparationEfficiency}
\end{figure}

Fig. \ref{fig:preparationEfficiency} a) shows the experimentally
determined preparation efficiency plotted against the angle between
the polarization of the light field near 369 nm and the direction of
the external magnetic field (with the detuning and intensity of the
light field near 369 nm already optimized. The solid line is meant
to guide the eye. In Fig. \ref{fig:preparationEfficiency} b) the
computed preparation efficiency is plotted for three different light
intensities (the Rabi frequency is given in units of the transition
linewidth). Numerical simulations show that the maximal preparation
efficiency achievable with a single light field near \nm{369} is
approximately 96.4\%. The experimentally obtained maximal value is
95.5 (6)\%.

Another approach to state preparation is resonant pumping of the
transition \one $\leftrightarrow$ \Pone by a light field near
\nm{369}. The short-lived \Pone state decays into both hyperfine
sublevels of the \Shalf\, ground state with a branching ratio of 2:1
It therefore takes only a few optical pumping cycles at a rate of
\MHz{19.6} to populate the \one state with close to 100\%
efficiency. This is much faster than the non-resonant depletion
process described above. Fig. \ref{fig:preparationEfficiency} c)
shows the computed preparation efficiency close to 100\% which, in
addition, is nearly independent of the polarization angle.

\ybodd possesses the simplest possible hyperfine structure of the
electronic ground state S$_{1/2}$,
%with states characterized by total angular momentum $F=0$ and $F=1$,
and allows for choosing either a magnetic field insensitive qubit or
qubit states whose energy separation depends on an applied magnetic
field. The latter choice is suitable for experiments where i) the
coupling between internal and external ionic degrees of freedom
(necessary for conditional quantum dynamics) relies on a state
dependent Zeeman force and ii) ions are individually addressed in
frequency space \cite{Mintert01}. In a suitably modified ion trap
microwave radiation may be used {\em directly} for coherent
manipulation of hyperfine qubits, thus eliminating all possible
sources of error that are present when first imprinting frequency
and phase information of microwaves onto laser radiation used for
driving a Raman transition, and then illuminating ions with this
light.

Financial support by the Deutsche Forschungsgemeinschaft, Science
Foundation Ireland under contract
No. 03/IN3/I397, and by the European Union (QGates) is
gratefully acknowledged. \\

\end{document}